\documentclass[12pt]{article}

\usepackage{graphicx}
\usepackage{scicite}
\usepackage{times}

\newcounter{lastnote}
\newenvironment{scilastnote}{%
\setcounter{lastnote}{\value{enumiv}}%
\addtocounter{lastnote}{+1}%
\begin{list}%
{\arabic{lastnote}.}
{\setlength{\leftmargin}{.22in}}
{\setlength{\labelsep}{.5em}}}
{\end{list}}

\title{Rotating Few-body Atomic Systems \\in the Fractional Quantum Hall Regime}
\author{Nathan Gemelke$^{1,2,\ast}$,Edina Sarajlic$^1$,Steven Chu$^3$\\
\\
\normalsize{$^1$Stanford University, Stanford, CA, USA 94305}\\
\normalsize{$^2$Department of Physics, University of Chicago, Chicago, IL, USA 60637}\\
\normalsize{$^3$U.S. Department of Energy, 1000 Independence Ave., SW Washington, DC, USA 20585}\\
\normalsize{$^\ast$To whom correspondence should be addressed; E-mail: ngemelke@uchicago.edu.}}
\date{}

\begin{document}
\baselineskip24pt
\maketitle
\begin{abstract}
Topologically-ordered matter is a novel quantum state of matter observed only in a small number of physical systems, notably two-dimensional electron systems exhibiting fractional quantum Hall effects \cite{tsui82}. It was recently proposed \cite{Cooper99,Wilkin2000,Cooper01} that a simple form of topological matter may be created in interacting systems of rotating ultra-cold atoms. We describe ensemble measurements on small, rotating clusters of interacting bosonic atoms, demonstrating that they can be induced into quantum ground states closely analogous to topological states of electronic systems. We report measurements of inter-particle correlations and momentum distributions of Bose gases in the fractional quantum Hall limit, making comparison to a full numerical simulation. The novel experimental apparatus necessary to produce and measure properties of these deeply entangled quantum states is described. 
\end{abstract}

The current understanding of many-body phases and phase transitions is dominated by the concept of spontaneously broken symmetry, as described by a locally-defined order parameter.  Topological matter is a fundamentally different type of ordered matter that cannot be characterized by a local order parameter and its long-range correlations \cite{Wen91}. Nor can transitions into these phases be described by Landau symmetry breaking, which until recently, was thought to describe \emph{all} phase transitions. In the quantum (zero-temperature) phase transition into these novel states of matter, particles form deeply entangled, long range states which can have unusual properties, such as excitations with fractional charge and fractional quantum statistics.  Some topological phases may also have applications in topological quantum computation \cite{Nayak08}.

In the fractional quantum Hall (FQH) effect, interacting electrons transition from a nearly free electron gas into a new strongly correlated state describable by composite particles of electrons associated with a number of magnetic flux quanta.  An analogous situation has been predicted to occur for an interacting Bose gas rotating a frequency $\Omega$ that approaches the harmonic oscillator frequency $\omega$ of the atoms confined in a trap. In this case, the quantized single-particle energy levels become highly degenerate, and the Hamiltonian of the system has a form similar to the interacting electron gas that produces fractional quantum Hall states. The lowest energy states transition from a weakly interacting Bose gas to highly entangled topological states at nonzero and quantized angular momentum.  Additionally, the excitations of this system are predicted to possess fractional statistical character \cite{Paredes2001}.

The atomic equivalent of a FQH state may be achieved only when the number of particles $N$ becomes comparable to the number of vortices $N_v$ in the ground state, i.e., when the filling factor $N / N_v$ is of order one.  Long-lived vortices have been excited in superfluid atomic gases in a variety of ways \cite{matthews99,Madison2000,haljan01,raman01,lin09}. In previous experiments probing high angular momentum Bose condensates consisting on the order of $10^4$ particles or more \cite{aboshaeer01,bretin04,Schweikhard2004}, the lowest Landau level was achieved, and a softening of the Abrikosov lattice of superfluid vortices was observed \cite{Schweikhard2004}.  However, in these experiments the FQH regime was not achieved, since the filling factor was of order 100, and the energetic gap to excited states, scaling inversely with $N$ \cite{CiracAdiabatic}, was significantly lower than achievable temperatures.

\begin{figure}[htp]\center
\includegraphics[width=4.0 in]{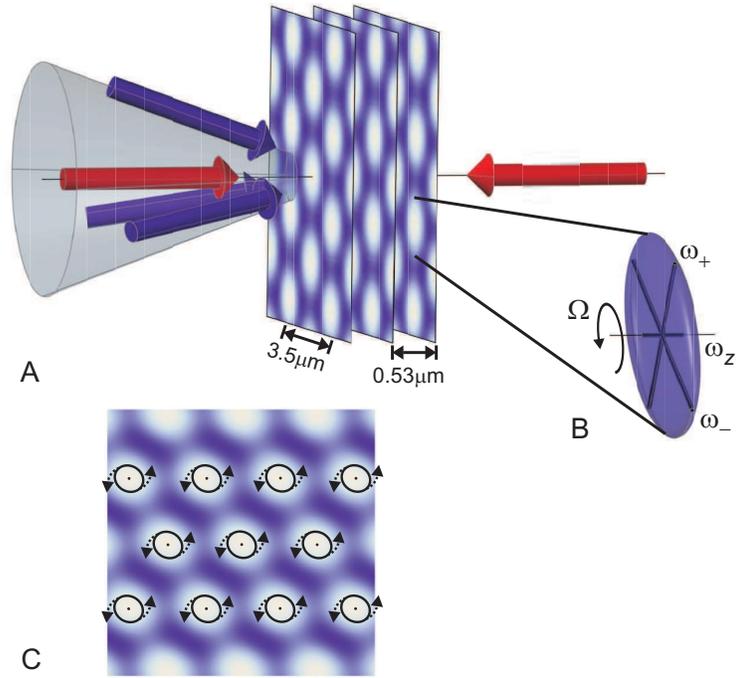}
\caption{ Simulation of a magnetic field at the rotating minima of an optical lattice potential. (A) Bose-condensed atoms are loaded into a deep optical lattice formed by the superposition of a two-dimensional lattice (blue beams) and a standing wave along its normal (red beams). The optical phases of the 2D lattice beams are modulated in time to produce lattice sites which near the lattice potential minima rotate at a rate comparable to the trapping frequency for atoms confined there. (B) The potential near each site approximates a strongly oblate harmonic oscillator, whose principle axes rotate in the plane normal to the direction of tightest confinement at a frequency $\Omega \sim \omega_\pm$  (C) Though the local potential near each site rotates, the overall lattice structure remains constant \cite{NDG_thesis}.} \label{fig1}
\end{figure}

In this article, we describe the realization of the FQH regime by probing an ensemble of interacting, rotating Bose gases in the few-body limit of $1 \leq N \leq 10$.  The key development in this work is the production of an optical lattice potential, each of whose sites exhibits a locally rotating deformation to its nominally isotropic and harmonic confining potential (see Fig. 1).  An adiabatic sequence of trap deformation strength and rotation rate is used to attempt to populate FQH ground states near the centrifugal limit. Following this manipulation, pulses of laser light which photo-associate pairs of atoms into electronically excited molecular states are used as a probe of inter-atomic correlation.  Additionally, time-of-flight images are taken to measure the momentum distribution of particles following the rotation sequence.  Both are shown to be in qualitative agreement with direct simulation of the few-body system, and provide evidence for strong correlations.

For an interacting gas of bosonic atoms confined to an oblate cylindrically-symmetric harmonic oscillator (see Fig. 1B, for $\omega_\pm=\omega$), the total energy in the rotating frame and the lowest Landau level may be written in units of the vibrational quanta $\hbar \omega$ as \cite{wilkin98}

\begin{equation}
H_{rot} = (1-\Omega/\omega)\hat{L}_{\Omega} + \eta \sum_{\{m_i\}} v_{\{m_i\} } \hat{a}^\dagger_{m_1} \hat{a}^\dagger_{m_2} \hat{a}_{m_3} \hat{a}_{m_4}
\end{equation}

where $\hat{L}_{\Omega}=\sum_m m \hat{a}^\dagger_{m}\hat{a}_{m}$ is the total angular momentum along the axis of rotation in units of $\hbar$, and $\hat{a}^\dagger_m$ creates a particle in an angular momentum eigenstate $m$ with wavefunction $\phi_m[z\!=\!(x\!+\!iy)/a_0]=z^m \exp(-|z|^2/2)$.  Here, $a_0=\sqrt{\hbar/m_a\omega}$, with $m_a$ the atomic mass. Excitation along the axis of rotation is effectively frozen out by the condition that the trapping frequency in that direction is much larger than the thermal and interaction energies.  Interactions are assumed repulsive and characterized by the dimensionless parameter $\eta = a_s / \sqrt{2\pi} a_\Omega$, where $a_s$ is the s-wave scattering length for the constituent atoms, $a_\Omega=\sqrt{\hbar/m_a\omega_z}$ is the confinement length along the rotation axis. The quantity $v_{\{m_i\}}$ characterizes the rate at which two atoms in angular momentum states $m_{1,2}$ are scattered into $m_{3,4}$ by interparticle interactions. Ground state correlation is expected to enter in the regime $\sqrt{2\pi}(N-1)\eta \sim (1-\Omega/\omega)$, where interactions are sufficient to begin mixing the single particle degeneracy near the centrifugal limit, and become strong in the limit $\sqrt{2\pi}\eta \sim (1-\Omega/\omega)$.  An energy spectrum for four particles is shown in Figure 2; the reduced interaction energy of strongly correlated states in general introduces level crossings between the zero angular momentum ground state and correlated states of higher $L_{\Omega}$ at rotation rates $\Omega<\omega$.

\begin{figure}[htp]\center
\includegraphics[width=4.25 in]{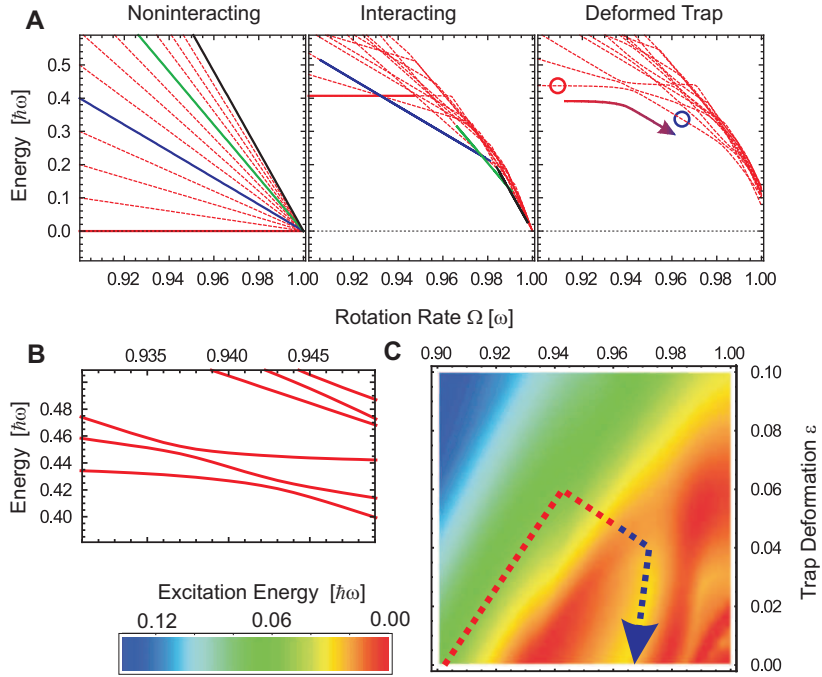}
\caption{ Energy spectrum and excitation gap as a function of trap rotation rate $\Omega$ and deformation strength $\epsilon$ for four particles.  (A) Energy spectrum for four particles in the lowest Landau level of a rotating harmonic trap.  Without interactions (left, $\eta = 0$), states collapse into a macroscopic degeneracy at the centrifugal limit $\Omega=\omega$.  States with total angular momentum in integer multiples of the particle number are shown as solid lines $L=0$ (red),$4$ (blue),$8$ (green),$12$ (black).  Repulsive interactions (middle, characterized by dimensionless parameter $\eta=0.0054$) shift the energy of uncorrelated states upward, perturb degeneracy, and introduce level crossings in the ground state contour toward increasingly correlated states. Adding a rotating quadrupole deformation (right, $\epsilon=0.016$) to the harmonic potential couples states whose total angular momentum differ by $2\hbar$ and produces avoided level crossings.  The ground state contour provides an adiabatic pathway into correlated states near the centrifugal limit with nonzero total angular momentum. (B) shows a small scale view of the the first ground-state crossing for four particles from a trivial non-rotating state to the four-particle Pfaffian with $L=4$. (C) shows the excitation gap as a function of both $\Omega$ and $\epsilon$.  The full adiabatic rotation sequence is illustrated as a dotted line, terminating at a final frequency $\Omega_f$; the entire sequence is translated in frequency as a control parameter.} \label{fig2}
\end{figure}

In this experiment, small numbers of atoms are first isolated and evaporatively cooled into the non-rotating ground states of many optical microtraps formed near the antinodes of an optical lattice.  Angular momentum is then transferred into the gases by applying a rotating deformation to the lattice potential near its antinodes, such that the \emph{local} potential approximates a rotating anisotropic harmonic oscillator.  This perturbation, describable as adding $H_\epsilon=\sum_{m}\epsilon\,(\hat{a}^\dagger_{m+2}\hat{a}_m+\hat{a}^\dagger_m\hat{a}_{m+2})$ to the energy $H_{rot}$ above, mixes states which differ by two units of angular momentum, and produces avoided crossings in the energy spectrum for the few-body states (Fig. 2b).  By controlling both the rotation rate $\Omega$ and deformation amplitude $\epsilon=2(\omega_+-\omega_-)/(\omega_++\omega_-)$, an adiabatic pathway is employed to transfer atom clusters from zero angular momentum into correlated states of higher angular momenta.  It is important to stress that rotation is applied only locally to each lattice site, and is not applied to the global lattice structure as presented elsewhere\cite{Cornell_mask}, allowing application of the rotating trap idea to small clusters of atoms.

At each ground state crossing, in the ideal case of a perfectly harmonic microtrap, $N\hbar$ units of angular momentum enter into a sample consisting of $N$ atoms, and simultaneously cause the second-order correlation of the gas to decrease by a significant degree.  This can be understood as the few-body equivalent of successive vortex nucleations, each accompanied by a strong redistribution of the few-body wavefunction away from the mean-field to accommodate energetic constraints of repulsion.   This continues as a sequence of tabulated \cite{Wilkin2000,Paredes2001,CiracAdiabatic} ground-state crossings terminating at a total angular momentum of $N(N-1)\hbar$, where the final state closely resembles\cite{Wilkin2000} the $\frac{1}{2}$-Laughlin $\phi({z_i})=\prod_{ij}(z_i-z_j)^2\exp(-\sum_i |z_i|^2)$, a bosonic analog of states \cite{laughlin} previously discussed in context of electronic systems, and where the probability to find two particles in the same location is zero.

As a specific example, for $N\!=\!4$ particles as shown in Fig. 2B, the ground-state transitions first from an uncorrelated $L\!=\!0$ state $\phi({z_i})=\prod_i\exp(-|z_i|^2)$, through a state of $L\!=\!4$ previously identified\cite{Wilkin2000,CiracAdiabatic} with a Pfaffian-type wavefunction\cite{Moore91}  at $\Omega=0.932\omega$.  The subsequent transition is to an $L\!=\!8$, $\frac{1}{2}$-quasiparticle\cite{CiracAdiabatic} state at $\Omega=0.977\omega$ before the final $L\!=\!12$, $\frac{1}{2}$-Laughlin state at $\Omega=0.989\omega$.

Atoms that adiabatically cross into the fractional quantum Hall ground states are expected to show a decrease in second-order correlation.  The degree of correlation in the few-body atomic states is probed with photoassociation to electronically excited molecules \cite{kinoshita05} (see methods).  The molecular formation rate $\Gamma\!\propto\!\int n^2(z)g_2(z)d^{\,2}\!z$, where the second-order correlation function $g_2(z)=\lim_{z' \rightarrow z }\langle \hat{\psi}^{\dagger}_z\hat{\psi}^{\dagger}_{z'}\hat{\psi}_{z'}\hat{\psi}_z\rangle/\langle \hat{\psi}^\dagger_z\hat{\psi}_z\rangle\langle \hat{\psi}^\dagger_{z'}\hat{\psi}_{z'}\rangle$ \cite{Naraschewski99}, with $\hat{\psi}_z=\sum_m \phi_m(z)\hat{a}_m$ the atomic field operator, measures the likelihood of pairs of atoms to be found at short range, compared to that expected from a mean-field form at the same density. The number of atoms surviving such a photoassociation pulse, therefore, directly probes the degree of correlation in the final few-body state of the atoms.

The optical lattice employed here (see Fig 1A and Methods) is the superposition of a two-dimensional triangular lattice potential formed by the intersection of three laser beams of equal frequency and a separate one-dimensional lattice formed by a separate counter-propagating beam pair along the rotation axis. The local lattice site structure approximates a strongly oblate, cylindrically symmetric three-dimensional harmonic oscillator potential with trapping frequencies of $(\omega_\pm,\omega_z)=2\pi\times(2.1,28)$kHz.  This results in relatively strong interactions between atoms, with interaction parameter $\eta=0.0054$, separately verified by measurement of collective mode frequencies (see Methods).  Care was taken to produce nominally symmetric lattice sites with $\omega_+\simeq\omega_-$ to an accuracy of 0.5\% over a day.  The method used to generate a controlled rotating trap deformation consists of diabatic translation of the two-dimensional lattice potential obtained by rapid electrooptic modulation of lattice beam phases, and is described in detail in the methods section below.  The time-average lattice potential has a static overall lattice structure, whose sites approximate locally harmonic oscillators with dynamically controlled anisotropy $\epsilon$ and orientation of their principle axes.

Atoms are loaded into this potential from a magnetically trapped Bose-condensate of $^{87}$Rb atoms, by first adiabatically increasing the two-dimensional lattice potential to confine atoms into tubes formed by the intersecting blue beams shown in Fig. 1A, then allowing the gas to expand along the axial direction. The axial standing wave formed by the red beams (Fig. 1A) is then adiabatically increased to a final depth sufficient to prevent tunnelling along the axial direction for relevant timescales.  This is followed by state-selective evaporation of atoms in lattice sites at the outer edges of the initial Bose Einstein condensate (see Methods), and selective evaporative removal of atoms to a final mean density of $\overline{N}\sim 5$ atoms/site in the full three dimensional potential. This results in a total number of atoms of order $10^4$, distributed over several thousand lattice sites.

The local lattice potential is then rotated with a time-dependent amplitude and frequency ramp determined to be a sufficiently adiabatic pathway to correlated states using a full numerical diagonalization of the hamiltonian in eq. 1, as described in the Methods section.  The few-body energy spectrum and gap to first excited state are presented in figure \ref{fig2} for $N=4$ atoms, along with the chosen ramp sequence (Fig. 2C).   The total ramp time used in this experiment is 48ms, consisting of three 16ms piecewise linear portions connecting the points $((\Omega-\Omega_f)/\omega,\epsilon)=(-0.065,0),(-0.025,0.06),(+0.005,0.04),(0,0)$.  The sequence was translated in frequency, to allow the final rotation rate $\Omega_f$ to be used as a control parameter. We note that the adiabatic ramp should serve to couple multiple occupancy classes into correlated states simultaneously, as each level of occupancy $N$ carries a similar, but different ground state crossing sequence.

\begin{figure}[htp]\center
\includegraphics[width=4.0 in]{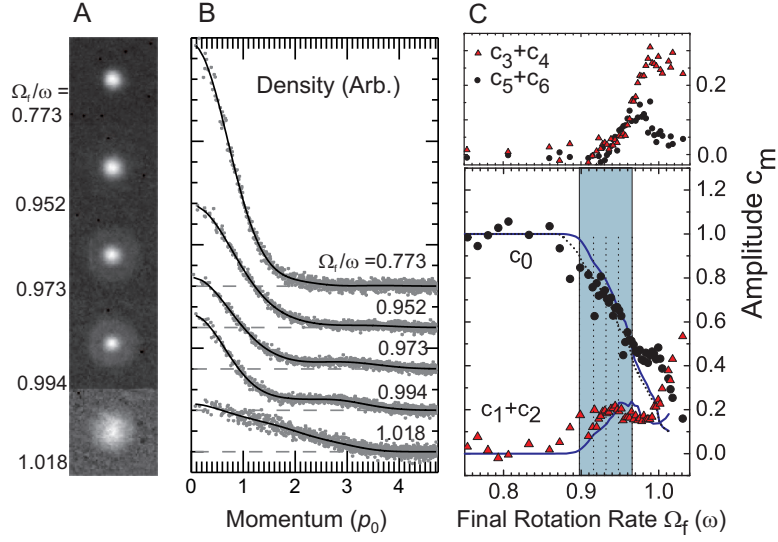}
\caption{ Angular momentum distribution of atoms in rotating lattice potential.  Absorbtion images (A) probe the angular momentum distribution following adiabatic rotation sequences at variable final rotation rates. As the rotation rate is increased, the cloud profile first grows slightly in radius, followed by the development of a halo of high angular momentum components.  Close to the centrifugal limit, the distribution broadens rapidly. Plotting density as a function of radius from the center-of-mass (B), the momentum distribution can be fit well to the form described in the text. (C) The expansion coefficients $c_m$ display a nontrivial behavior as the centrifugal limit is approached from below.  The dependence of the lowest angular momentum components $c_0$ and $c_1+c_2$ are compared to numeric predictions (blue lines).  In the blue region, marking the range spanned by the first ground-state crossing location for particle numbers $2 \leq N \leq 6$ (vertical lines), we find reasonable quantitative agreement with the model, implying successful generation of FQH states of angular momentum $N$.  The small discrepancy seen at rotation rates $\Omega_f \sim 0.88$ are expected from a truncation of numeric predictions at occupancies $N \geq 7$; incorporating a phenomenological prediction (dotted curve, see Methods) for $N=7$ improves the fit.} \label{fig3}
\end{figure}

The final state is probed by simply releasing the atoms from all applied potentials. The resulting atomic distribution is then imaged along the rotation axis after a $10$ms time-of-flight.  Due to the tight confinement of the lattice, the evolution in time-of-flight is primarily determined by the initial momentum distribution before release and is largely unperturbed by interactions. Furthermore, due to extremely weak tunnel coupling between sites, the time-of-flight density is an incoherent sum of the momentum distributions in all sites. In the lowest Landau level, time-of-flight expansion proceeds such that the momentum distribution mimics the in-situ density $n(z)$ with $|z|$ replaced by $p/p_0$\cite{read03}, where $p$ is the magnitude of momentum in the plane of rotation, and $p_0=(m_a\hbar\omega)^{1/2}$ is the oscillator momentum. This leads to a momentum space distribution of the form $n(p)=\sum_m c_m (p/p_0)^{2m} \exp(-p^2/p_0^2)/m!$, where $c_m$ for $m=0,1,2...$ characterize the angular momentum distribution.

At sufficiently low final rotation rates, the adiabatic pathway terminates prior to the first ground state crossing, and the momentum distribution should reflect the zero angular momentum ground state.  For rotation rates $\Omega_f<0.87\omega$, we observe a gaussian time-of-flight profile of momentum width $m_a\times3.8$mm/s, consistent with atoms remaining in the ground state, given $p_0\sim m_a\times3.1$mm/s, the imaging resolution and initial spatial extent of atoms across the lattice. This verifies an ability to provide controlled rotation sequences without unanticipated heating.  The angular momentum distribution $c_m$ can be extracted from images using singular value decomposition to the fit form for $n(p)$ above, and as described in the methods; the result is shown in figure 3C. At rotation rates $0.87<\Omega_f/\omega<0.94$, the gaussian distribution broadens, indicating increased occupancy of single-particle states of angular momentum $m=1$.  For $\Omega_f \ge 0.95\omega$, a high momentum "halo" develops with significant population in $m\geq 6$, growing in amplitude for frequencies $0.950<\Omega_f/\omega<0.980$. These halos are unexpected in the idealized case of a perfectly harmonic oscillator potential, but appear in the coherent evolution of higher occupancy clouds for an anharmonic rotating trap with a negative quartic component, as described in the methods below.  Beyond a final rotation rate $\Omega_f > 1.01\omega$, the central peak in time-of-flight grows in size rapidly, reflected by a growth in components $c_m$ for $1 \ge m \le 4$.

The expected form of $n_N(p)$ was calculated from full dynamic evolution of the hamiltonian presented above for all occupancies $N<7$ as described in the Methods.  A weighted sum of the $n_N(p)$ was taken to represent the expected distribution of occupancies in the lattice, and is shown in figure 3C, compared to experimentally recorded profiles.  In general, one finds reasonable quantitative agreement between model and experiment over the range $0.91<\Omega_f<0.97$; the behavior in this range is expected from numerics (see methods) to be dominated by the first ground-state level crossing of successively lower $N$ systems as $\Omega_f$ increases.  Each of these crossings is expected to introduce significant correlation, and its precise location is determined by the degree to which the repulsive interaction energy is reduced by correlations. For $N=4$ particles, this level crossing has been associated with the entrance of a few-body Pfaffian ground state \cite{Wilkin2000,CiracAdiabatic}.  The disagreement at $0.86<\Omega_f<0.90$ and slight mismatch in slopes for $0.91<\Omega_f<0.97$ is understandable, due to a truncation of the numeric predictions for $N>6$ where calculations become too numerically intensive; this can be improved assuming a phenomenological form for N=7 as described in Figure \ref{fig3} and methods.  The upturn in the data very close to the centrifugal limit $\Omega_f>0.97$ is currently not understood, but occurs where the energy spectrum is exceedingly complicated and sensitively dependent on trap shape.

To probe atom-atom correlation in the final state, pulses of light tuned to a photoassociation transition were applied immediately following the adiabatic rotation sequence. This transfers pairs of atoms in close proximity into electronically excited molecules, which quickly relax to deeply bound states with sufficient excess energy to be ejected from the trap. The remaining atoms were released within $100\mu$s and imaged in time-of-flight. The total surviving atom number $N_{pa}$ then reflects the probability to find atoms in the ground state at short interatomic range.  Details of this technique appear in the methods and reference \cite{kinoshita05}.

\begin{figure}[htp]\center
\includegraphics[width=4.0 in]{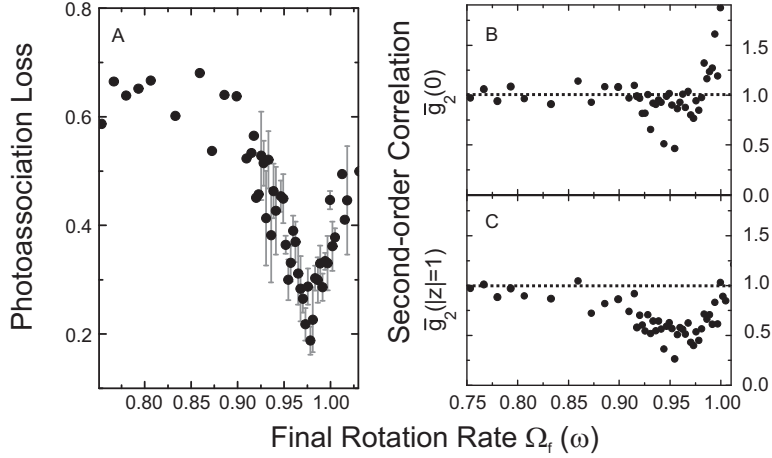}
\caption{Probing atomic correlation with photoassociation. (A) Shows fraction $(N_{npa}-N_{pa})/N_{npa}$ of atoms lost after rotation sequence and short pulse of photoassociation light.  The strong suppression at the centrifugal limit is indicative of strong correlation. (B,C) Measurement of density profile with and without photoassociation allows extraction of the second-order correlation function $g_2(z)$ as defined in the text.  The depression near the centrifugal limit is most apparent at nonzero radius.} \label{fig1}
\end{figure}

From Fig. 4, one can see a strong suppression of photoassociation loss near the centrifugal limit, consistent with the introduction of correlation in the few-body wavefunction. Here we plot the fraction of atoms lost to photoassociation as $(N_{npa}-N_{pa})/N_{npa}$, where $N_{pa}$($N_{npa}$) is the number measured with (without) the photo-association laser pulse. To establish that the change in loss rate is due to change in correlation, and not simply decrease in overall density, we form a simplistic estimate of the second-order correlation function $g_2(z,\Omega_f)$ as defined above.  Since one expects a constant $g_2(z,\Omega_f)=(N-1)/N$ for any $N$-particle mean-field state of the form $(\sum_m b_m \hat{a}^\dagger_m)^N|0\rangle$, independent of the particular state defined by the $b_m$, a reduction of $g_2(z, \Omega_f)$ near the centrifugal limit would provide evidence for departure from the mean-field into the fractional quantum Hall regime.  We model the loss rate as $\dot{n_0}(z,t)=-Kg_2(z,\Omega_f) n_0^2(z,t)$, where $n_0$ is the atomic density on a given site, $K$ is a rate constant, and $t$ is the pulse time. Using this, an estimate of $g_2$ can be formed as $g_2(z,\Omega_f)\sim [n_0^{-1}(z,\tau)-n_0^{-1}(z,0)](K\tau)^{-1}$.  The quantity $K\tau$ is eliminated by forming $\bar{g}_2(z)=g_2(z,\Omega_f)/g_2(0,0)$, where $g_2(0,0)$ is estimated by averaging $g_2(0,\Omega_f)$ for all points with $\Omega_f< 0.875\omega$; we find the results plotted in Figs. 4B and 4C.

In this analysis, we have treated the density after time-of-flight as if it arose from a well defined occupancy in the lattice; in reality the time-of-flight density should be treated as an incoherent sum over the density $n_N(p)$ arising at different occupancies $N$, and the estimate $g_2(z)$ as a type of occupation-averaged correlation.  The presence of systems with lower particle number contributes a background of uncorrelated low angular momentum states near the centrifugal limit, making the reduction in $g_2(z)$ there most apparent at nonzero radius.  A more sophisticated model could give a more realistic interpretation of the quantity $g_2(z)$ in terms of a weighted average of the second-order correlation of each occupancy class. By systematic variation of the average occupancy in the lattice, one could hope to extract separately the correlation function for each occupancy.

In summary, an apparatus has been described to bring a quantum degenerate system of bosonic atoms into the quantum-Hall regime by applying a precisely controlled trap rotation.  We have designed and implemented an adiabatic sequence of trap deformations to populate highly correlated states of strongly interacting atoms at nonzero angular momentum, and probed its efficiency by measurement of both momentum distributions and interatomic correlation using photoassociation techniques. Quantitative agreement between experimental observations and full numeric models indicates the establishment of a viable platform to further study entangled atomic states in the fractional quantum Hall regime.

\bibliography{FQHall_ver5}

\begin{thebibliography}{10}

\bibitem{tsui82}
D.~C. Tsui, H.~L. Stormer, A.~C. Gossard, {\it Phys. Rev. Lett.\/} {\bf 48},
  1559 (1982).

\bibitem{Cooper99}
N.~R. Cooper, N.~K. Wilkin, {\it Phys. Rev. B\/} {\bf 60}, R16279 (1999).

\bibitem{Wilkin2000}
N.~K. Wilkin, J.~M.~F. Gunn, {\it Phys. Rev. Lett.\/} {\bf 84}, 6 (2000).

\bibitem{Cooper01}
N.~R. Cooper, N.~K. Wilkin, J.~M.~F. Gunn, {\it Phys. Rev. Lett.\/} {\bf 87},
  120405 (2001).

\bibitem{Wen91}
X.~G. Wen, {\it Int. J. Mod. Phys. B\/} {\bf 5}, 1641 (1991).

\bibitem{Nayak08}
C.~Nayak, S.~H. Simon, A.~Stern, M.~Freedman, S.~Das~Sarma, {\it Rev. Mod.
  Phys.\/} {\bf 80}, 1083 (2008).

\bibitem{Paredes2001}
B.~Paredes, P.~Fedichev, J.~I. Cirac, P.~Zoller, {\it Phys. Rev. Lett.\/} {\bf
  87}, 010402 (2001).

\bibitem{matthews99}
M.~R. Matthews, {\it et~al.\/}, {\it Phys. Rev. Lett.\/} {\bf 83}, 2498 (1999).

\bibitem{Madison2000}
K.~W. Madison, F.~Chevy, W.~Wohlleben, J.~Dalibard, {\it Phys. Rev. Lett.\/}
  {\bf 84}, 806 (2000).

\bibitem{haljan01}
P.~C. Haljan, I.~Coddington, P.~Engels, E.~A. Cornell, {\it Phys. Rev. Lett.\/}
  {\bf 87}, 210403 (2001).

\bibitem{raman01}
C.~Raman, J.~R. Abo-Shaeer, J.~M. Vogels, K.~Xu, W.~Ketterle, {\it Phys. Rev.
  Lett.\/} {\bf 87} (2001).

\bibitem{lin09}
Y.~J. Lin, R.~L. Compton, K.~Jimenez-Garcia, J.~V. Porto, I.~B. Spielman, {\it
  Nature\/} {\bf 462}, 628 (2009).

\bibitem{aboshaeer01}
J.~R. Abo-Shaeer, C.~Raman, J.~M. Vogels, W.~Ketterle, {\it Science\/} {\bf
  292}, 476 (2001).

\bibitem{bretin04}
V.~Bretin, S.~Stock, Y.~Seurin, J.~Dalibard, {\it Phys. Rev. Lett.\/} {\bf 92}
  (2004).

\bibitem{Schweikhard2004}
V.~Schweikhard, I.~Coddington, P.~Engels, V.~P. Mogendorff, E.~A. Cornell, {\it
  Phys. Rev. Lett.\/} {\bf 92}, 040404 (2004).

\bibitem{CiracAdiabatic}
M.~Popp, B.~Paredes, J.~I. Cirac, {\it Phys. Rev. A\/} {\bf 70}, 053612 (2004).

\bibitem{NDG_thesis}
N.~Gemelke, Quantum degenerate gases in controlled optical lattice potentials,
  Ph.D. thesis, Stanford University (2007).

\bibitem{wilkin98}
N.~K. Wilkin, J.~M.~F. Gunn, R.~A. Smith, {\it Phys. Rev. Lett.\/} {\bf 80},
  2265 (1998).

\bibitem{Cornell_mask}
S.~Tung, V.~Schweikhard, E.~A. Cornell, {\it Phys. Rev. Lett.\/} {\bf 97}
  (2006).

\bibitem{laughlin}
R.~B. Laughlin, {\it Phys. Rev. Lett.\/} {\bf 50}, 1395 (1983).

\bibitem{Moore91}
G.~Moore, N.~Read, {\it Nucl. Phys. B\/} {\bf 360}, 362 (1991).

\bibitem{kinoshita05}
T.~Kinoshita, T.~Wenger, D.~S. Weiss, {\it Phys. Rev. Lett.\/} {\bf 95}, 190406
  (2005).

\bibitem{Naraschewski99}
M.~Naraschewski, R.~J. Glauber, {\it Phys. Rev. A\/} {\bf 59}, 4595 (1999).

\bibitem{read03}
N.~Read, N.~R. Cooper, {\it Phys. Rev. A\/} {\bf 68}, 35601 (2003).

\bibitem{Fioretti01}
A.~Fioretti, {\it et~al.\/}, {\it Eur. Phys. J. D\/} {\bf 15}, 189 (2001).

\end{thebibliography}
\bibliographystyle{Science}

\begin{scilastnote}
\item This work was funded by in part by grants from the NSF and AFOSR. The authors wish to thank B. Paredes, E. Mueller, C. Wu and C. Chin for insightful comments.  NDG was supported in part by the Grainger foundation.
\end{scilastnote}
\clearpage

\section{Methods}

\subsection{Rotating Lattice Production}
To produce an optical lattice of locally rotating potentials, three laser beams of equal intensity and detuned far from atomic resonance are combined with their propagation directions evenly distributed on the surface of a cone with a small apex angle.  The optical interference pattern created by these beams consists of a triangular lattice of intensity maxima, whose light shift form a conservative trapping potential for atomic motion. Near to the minima, the potential is locally harmonic, approximately cylindrically symmetric, and may be described by $V(x,y)=-V_0\sum_j \cos(\sqrt{3}k_e r_j+\phi_j)$, where $r_j=\cos(2\pi j/3)x+\sin(2\pi j/3)y$, with $x,y$ coordinates in the lattice plane, $k_e=2\pi\sin\theta/\lambda$, and $\phi_j$ represent the relative optical phases of the three beams (Fig. 1A). By choosing a small intersection angle $\theta=8$ degrees, the spacing between lattice sites is 3.5$\mu m$. This reduces the tunneling rate of atoms between lattice sites in the 2D potential to be negligible on experiment timescales, and simultaneously makes the potential effectively more harmonic by reducing its vibration frequency at a fixed total depth.  The three lattice beams originate from a common $1.5$W, fiber-coupled beam, intensity stablized by an acousto-optic modulator and derived from a $10$W single-mode Nd:YAG ring laser injection-locked to a stable $0.5$W non-planar ring oscillator (Lightwave NPRO).  To produce a nominally cylindrically-symmetric potential near the bottom of each lattice site, center-of-mass vibration frequencies are measured with an accuracy of 0.3\% along two directions, and beam intensities adjusted to equalize these frequencies to a typical precision of 0.5\% over a typical experiment run time of several hours.  The inhomogeneous spread of vibration frequencies due to the gaussian intensity profile of the lattice beams is calculated to be 0.3\% over the loaded volume.

The locally rotating potential wells are created by inserting two electrooptic phase modulators into two of the beams forming the 2D lattice potential.  Manipulating the relative phases $\phi_j$ of the beams causes a 2D translation of the lattice potential.  By rapidly scanning the potential along a given direction $\phi_j=\Delta\phi \cos(\theta-2\pi j /3)\sin(\omega_{RF} t)$ at a frequency $\omega_{RF}\gg\omega,\omega_z$, the lattice potential may be effectively averaged over the timescale for atomic motion creating a local potential with a reduced vibration frequency along the axis of translation (defined by $\theta$).  By slowly pivoting this axis in time according to $\theta=\Omega t$ at a rate of order $\Omega\sim O(\omega_r)$, the time-averaged local potential approximates an anisotropic harmonic oscillator whose principle axes rotate at rates which can be comparable to the trapping frequency.  For the data shown here, $\omega_{RF}=2\pi\times 500$kHz, which was separately verified to be sufficiently diabatic that heating did not occur on relevant timescales by measuring cloud width in time-of-flight.  Averaging over the short timescale, the lattice potential can then be written as $V(x,y)=-\sum_j V_j \cos(\sqrt{3}k_e r_j)$, where $V_j=V_0 J_0(\Delta\phi/2\sin(\Omega t +2\pi j/3))$, and $J_0$ is the zeroth-order Bessel function.  The local potential near each site minimum is then approximately (dropping a constant) $V(r,\phi) \sim ( m\omega^2 r^2 / 2)(1+2\epsilon\cos(2\phi))$ in cylindrical coordinates $(r, \phi)$, with $\omega=\omega_0(1-\Delta\phi^2/32)$ and $\epsilon=\Delta\phi^2/64$, with $\omega_0$ the radial vibration frequency of a lattice site without modulation.  The adiabatic pathway described in the text is achieved by controlling the amplitude $\Delta\phi$ of RF phase modulation and the pivot frequency $\Omega$.

An additional 1D potential is used to isolate small clusters of atoms and enhance the effect of interactions between atoms loaded into this potential using two additional beams, frequency-offset from those forming the 2D potential and counterpropagating along the rotation axis.  The final potential is a three-dimensional array of highly oblate three dimensional harmonic oscillators, with radial trapping frequencies of up to 6kHz and axial frequencies up to 30kHz. Interaction strength is characterized by the ratio of scattering length to oscillator length from confinement along the rotation direction, $\eta=a_s/\sqrt{2\pi} a_\Omega\sim0.0054$.  The parameter $\eta$ was separately checked by measuring the variation of the collective quadrupole mode frequency with mean particle number \emph{(M1)}. over the range 3-100 atoms/site, showing a characteristic curve from the hydrodynamic to single-particle regime consistent with the expected interaction strength and separately measured center-of-mass frequencies.

\subsection{Preparation of Initial State}
Atoms are loaded from a $^{87}$Rb Bose-Einstein condensate of $10^5$ particles in the $|F=2, m_F=2\rangle$ state at a temperature of roughly $30nK$ formed by evaporative cooling in a time-orbiting-potential (TOP) magnetic trap.  After evaporation, the two-dimensional lattice potential is adiabatically increased from zero intensity to its full value of $500mW$ per beam, after which the TOP trap is deformed into a quadrupole trap whose center is pulled below the position of the atoms loaded into the tubelike 2D lattice potential.  The axial confinement of atoms trapped in the two-dimensional lattice potential is adiabatically decreased by reducing the magnetic quadrupole field, during which time the axial size of the cloud increases from an a Thomas-Fermi radius of 20$\mu $m to a half-width of approximately 200$\mu$m.  Following this, the axial standing wave intensity is increased slowly to a depth sufficient to inhibit axial motion. In order to produce a more defined mean occupancy in the full three-dimensional lattice potential, a tomographic technique is used to remove atoms in the tail of the density distribution along the axial direction. A weak magnetic field gradient is applied, and a microwave field is applied to transfer atoms from the $|F=2, m_F=2\rangle$ state into $|F=1, m_F=1\rangle$, whose magnetic moment is opposite in sign.  By slowly sweeping the microwave frequency, atoms are adiabatically transferred between internal states at the edges of the cloud.  Following this, the two-dimensional lattice depth is reduced, and a strong magnetic field gradient is applied to completely remove atoms in the state $|F=1, m_F=1\rangle$ while slowly evaporating atoms in $|F=2, m_F=2\rangle$ until the desired mean occupancy is reached, as inferred from absorbtion imaging performed transverse to the rotation axis. The density profile along the axial direction after this process resembles a top-hat, with a physical width of 150$\mu$m.  Following this, the depth of the two-dimensional lattice is increased to its full value.

\subsection{Analysis of Time-of-Flight}
Time-of-flight momentum distributions were obtained by releasing atoms from all applied potentials suddenly, allowing free expansion of the atoms for 10ms, then performing absorption imaging along the rotation axis.  The flight duration was chosen as a balance to obtain a dominantly momentum-space picture of the cloud, while preserving sufficient density for acceptable signal to noise.  Profiles were fit to the form $n(p)=\sum_m c_m (p/p_0)^{2m} \exp(-p^2/p_0^2)/m!$ for the coefficients $c_m$ using a singular value decomposition.  The value of $p_0$ was extracted from data taken at small $\Omega_f$, and fixed for remaining data.  A constraint of positivity was enforced on the coefficients $c_m$ to remove small negative values by minimizing $\sum_{mm'}\delta c_m C_{mm'}\delta c_{m'}+\sum_m\lambda(c_m+\delta c_m)^2$ with $\lambda=1$, where $C_{mm'}$ is the covariance matrix obtained from singular value decomposition.

\subsection{Photoassociation-based Probe of Short-range Correlation}
Following the adiabatic evolution sequence, correlation is probed by driving pairs of colliding atoms into bound, electronically-excited molecules with a short pulse of light tuned from the $|2,2\rangle$ free-atom continuum to the $|0_g^-,\nu=1,J=2\rangle$ molecular state\cite{Fioretti01,kinoshita05} at 12,789 cm$^{-1}$. Light was derived from a external-cavity-diode laser frequency stabilized to $\sim7$MHz to an optical transition in molecular iodine.  After photoassociation, rapid relaxation to the ground state is assumed to be accompanied by sufficient kinetic energy that the resultant atoms are lost from the lattice with a high probability, and hence the loss rate from the trap during the pulse reflects the probability to find two atoms at short range (determined by the outer classical turning point of the excited molecular state at $r_0\sim2$nm \cite{Fioretti01}).  Furthermore, the pulse time is made short (100$\mu$s), such that no significant dynamic evolution of the few body state occurs during the exposure to photoassociation light aside from loss.  After the photoassociation pulse, atoms are released in time-of-flight, and absorptively imaged to obtain both their total number and momentum distribution.  The time-of-flight image is recorded in successive experiments both with and without the photoassociation strobe but the same adiabatic preparation sequence, and compared.  Total atom number (presented in figure 4A) is extracted by direct integration of the raw absorbtion images without additional fitting or modeling.  To extract the estimate of second-order correlation, the density distribution before and after time-of-flight is decomposed according to the fitting method described above.

\subsection{Modeling Adiabatic Evolution}
To model the few-body states, their dynamic evolution, and degree of correlation, direct diagonlization of the few-body hamiltonian $H_{rot}$ (eq. 1 in the text) including the effects of the rotating trap and short-ranged repulsive interactions was performed, similar to that presented in Ref. \emph{16}.  Interactions were modeled by a contact potential, leading to the form of the scattering matrix $v_{\{ m_i \}}=\delta_{m_1+m_2}^{m_3+m4}(m_1+m_2)!\,2^{m_1+m_2}(\prod_i m_i!)^{-1/2}$, where $\delta_i^j$ is a kroneckar delta.  Direct diagonalization for fixed $\epsilon$ and $\Omega$ were used to guide the choice of adiabatic evolution in trap rotation rate and deformation strength presented in figure 2.  The full evolution of the few body states, using a time-dependent Shrodinger's equation with varying $\epsilon(t)$ and $\Omega(t)$, was used to calculate the expected momentum distribution presented in figure 3C, including effects of nonadiabaticity.  The effects of trap anharmonicity on the spectrum and dynamic evolution were included by modifying the single-particle energy levels to first order according to the expected form of the lattice potential at finite depth.

\begin{figure}[htp]
\vspace{.25in}\center
\includegraphics[width=4.0 in]{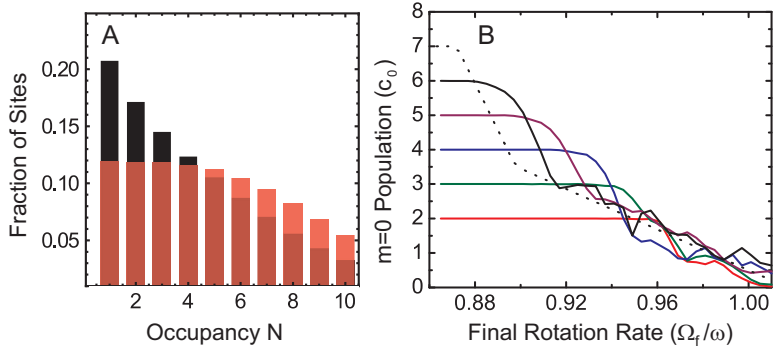}
\caption{Modeling time-of-flight signatures.  (A) The expected distribution $P_N$ of lattice site occupancy $N$ across the lattice volume with (without) tomographic selection is shown in red (black).  (B) A separate numeric calculation for the full few-body quantum state is performed for each occupancy $N$; here the lowest angular momentum component $c_{m=0}$ is shown for occupancies $2\leq N \leq 6$. Each occupancy displays a large drop at the position of the first ground state level crossing, followed by a steady redistribution of atoms to higher angular momentum. For $N\geq 7$, numeric calculation is too intensive, and a phenomenological form is assumed (dashed line).  Weighting the results in (B) by the distribution in (A), and accounting for effects of imaging resolution and cloud size, we obtain the curve presented in Figure 3C of the main text.  Calculations were performed for the adiabatic ramp parameters described in the text, and $\eta=0.0054$. } \label{fig1}
\end{figure}

To account for multiple occupancy classes in the dynamic evolution of the angular momentum distribution, a separate few-body numeric evolution was performed for each occupancy class $N$, (results are shown in Figure 5), and a weighted sum of the angular momentum distribution was taken for the results shown in Figure 3C of the main text.  To approximate the distribution of occupancies $N$, the number of lattice sites containing $N$ atoms was taken to be Poisson $P_N(\nu)=\nu^N\exp(-\nu)/N!$ with mean number $\nu$. While it is in principle possible the preparation sequence favors a sub-Poisson distribution through interaction-driven localization, we assume this (at finite temperature and over the range of filling factors produced) to be a small effect on the full distribution $P_N$. The mean $\nu$ was assumed to vary spatially across the cloud, following a Thomas-Fermi profile in the radial direction, and approximately flat along the axial direction due to the tomographic selection described above.  For an inverse quadratic density profile of maximum filling $\nu_0$, the distribution then assumes the form $P_N \propto \int_0^{\nu_0}d\nu\, \nu^N\exp(-\nu)/N!$.  The peak filling $\nu_0\sim10$ was estimated from absorbtion imaging.  The effect of finite imaging resolution and initial cloud size is accounted for by assuming the time-of-flight profile suffers a gaussian blur of $1/e$-radius $\mathcal{R}\sim0.7p_0$ estimated from cloud profiles measured at low $\Omega_f$, and adjusting the numeric prediction of the density distribution presented in Figure 3C of the text according to $c_{m=0} \rightarrow \sum_m c_m \alpha_m / \sum _m \alpha_m$, where $\alpha_m^{-1}=(p_0^{-2}+\mathcal{R}^{-2})^{m+1}p_0^{2m}\mathcal{R}^{2}$.
\end{document}